\input amstex
\documentstyle{amsppt}
%
\catcode`@=11
\redefine\output@{%
  \def\break{\penalty-\@M}\let\par\endgraf
  \ifodd\pageno\global\hoffset=105pt\else\global\hoffset=8pt\fi  
  \shipout\vbox{%
    \ifplain@
      \let\makeheadline\relax \let\makefootline\relax
    \else
      \iffirstpage@ \global\firstpage@false
        \let\rightheadline\frheadline
        \let\leftheadline\flheadline
      \else
        \ifrunheads@ 
        \else \let\makeheadline\relax
        \fi
      \fi
    \fi
    \makeheadline \pagebody \makefootline}%
  \advancepageno \ifnum\outputpenalty>-\@MM\else\dosupereject\fi
}
\def\Beta{\mathchar"0\hexnumber@\rmfam 42}
\catcode`\@=\active
\nopagenumbers
\chardef\textvolna='176

\chardef\bigalpha='013
\def\negskp{\hskip -2pt}
\def\divr{\operatorname{div}}
\def\rot{\operatorname{rot}}

\def\const{\operatorname{const}}

\def\compos{\,\raise 1pt\hbox{$\sssize\circ$} \,}
\def\blue#1{#1}

\catcode`#=11\def\diez{#}\catcode`#=6
\catcode`&=11\catcode`&=4
\catcode`_=11\def\podcherkivanie{_}\catcode`_=8
\catcode`~=11\catcode`~=\active
\def\mycite#1{\cite{\blue{#1}}\immediate\special{ps:
     ShrHPSdict begin /ShrBORDERthickness 0 def}}

\def\mytag#1{%
    \tag#1}
\def\mythetag#1{\thetag{\blue{#1}}\immediate\special{ps:
     ShrHPSdict begin /ShrBORDERthickness 0 def}}
\def\myrefno#1{\no#1}
\def\myhref#1#2{\blue{#2}\immediate\special{ps:
     ShrHPSdict begin /ShrBORDERthickness 0 def}}
\def\myEarXivlink{\myhref{http://arXiv.org}{http:/\negskp/arXiv.org}}

\def\mytheorem#1{\csname proclaim\endcsname{Theorem #1}}
\def\mytheoremwithtitle#1#2{\csname proclaim\endcsname{Theorem #1#2}}

\def\mylemma#1{\csname proclaim\endcsname{Lemma #1}}
\def\mylemmawithtitle#1#2{\csname proclaim\endcsname{Lemma #1#2}}
\def\mythelemma#1{\blue{#1}\immediate\special{ps:
     ShrHPSdict begin /ShrBORDERthickness 0 def}}
\def\mycorollary#1{\csname proclaim\endcsname{Corollary #1}}

\def\mydefinition#1{\definition{Definition #1}}

\def\myconjecture#1{\csname proclaim\endcsname{Conjecture #1}}
\def\myconjecturewithtitle#1#2{\csname proclaim\endcsname{Conjecture #1#2}}

\def\myproblem#1{\csname proclaim\endcsname{Problem #1}}
\def\myproblemwithtitle#1#2{\csname proclaim\endcsname{Problem #1#2}}

\pagewidth{360pt}
\pageheight{606pt}
\topmatter
\title
A model with two quantum particles 
similar to the hydrogen atom.
\endtitle
\rightheadtext{A model with two quantum particles \dots}
\author
Ruslan Sharipov
\endauthor
\address Bashkir State University, 32 Zaki Validi street, 450074 Ufa, Russia
\endaddress
\email\myhref{mailto:r-sharipov\@mail.ru}{r-sharipov\@mail.ru}
\endemail
\abstract
     The hydrogen atom with the Coulomb interaction is one of the exactly 
solvable non-relativistic quantum models. Unlike many other exactly solvable 
models it describes a real physical object providing the formulas for energy 
levels and stationary state wave functions of a real hydrogen atom. In this 
paper we modify the model replacing the Coulomb interaction by the interaction 
of the proton and the electron with the classical electromagnetic field 
serving as an intermediary transmitting the electromagnetic interaction 
of these two charged quantum particles. 
\endabstract
\endtopmatter
\TagsOnRight
\document

\head
1. Introduction.
\endhead
     The standard hydrogen atom is a system composed by the proton $p^{\sssize\,+}$ 
with the mass $m_p$ and the electron $e^{\sssize -}$ with the mass $m_e$. This system
is described by the Shr\"odinger equation $H\,\Psi=\Cal E\,\Psi$ with the following 
Hamilton operator:
$$
\hskip -2em
H=-\frac{\hbar^2}{2\,m_p}\,\nabla_{\kern -1pt p}^{\kern 1pt 2}
-\frac{\hbar^2}{2\,m_e}\,\nabla_{\kern -1pt e}^{\kern 1pt 2}
-\frac{e^{\kern 1pt 2}}{|\bold r_p-\bold r_e|} 
\mytag{1.1}
$$
(see \mycite{1} or \mycite{2}). In order to solve the Shr\"odinger equation 
$H\,\Psi=\Cal E\,\Psi$ the so called center-of-mass coordinates are used:
$$
\xalignat 2
&\hskip -2em
\bold R=\frac{m_e\,\bold r_e+m_p\,\bold r_p}{m_p+m_e},
&&\bold r=\bold r_e-\bold r_p.
\mytag{1.2}
\endxalignat
$$
Applying \mythetag{1.2} to \mythetag{1.1}, one can easily derive the following 
expression for $H$:
$$
\hskip -2em
H=-\frac{\hbar^2}{2\,(m_p+m_e)}\,\nabla_{\kern -1pt\bold R}^{\kern 1pt 2}
-\frac{\hbar^2}{2\,\tilde m}\,\nabla_{\kern -1pt\bold r}^{\kern 1pt 2}
-\frac{e^{\kern 1pt 2}}{|\bold r|} 
\text{, \ where \ }\tilde m=\frac{m_e\,m_p}{m_e+m_p}. 
\mytag{1.3}
$$
The formula \mythetag{1.3} then is used for separating variables in the Shr\"odinger 
equation $H\,\Psi=\Cal E\,\Psi$ and deriving the standard formulas for $\Cal E$ and 
$\Psi$.\par
      In this paper we choose a different approach by omitting the Coulomb interaction 
term in \mythetag{1.1} and writing the Hamilton operator in the following form:
$$
\hskip -2em
H=-\frac{\hbar^2}{2\,m_p}\,\nabla_{\kern -1pt p}^{\kern 1pt 2}
-\frac{\hbar^2}{2\,m_e}\,\nabla_{\kern -1pt e}^{\kern 1pt 2}
+H_{\text{int}}.
\mytag{1.4}
$$
Here $H_{\text{int}}$ is the {\bf interaction} term describing the \pagebreak
interaction of the classical electromagnetic field with the proton and the electron. 
The explicit expression for $H$ including the interaction term $H_{\text{int}}$ can 
be obtained from the formula \mythetag{3.6} below.\par
     Apart from \mythetag{1.4}, we consider the backward influence of two quantum
particles upon the classical electromagnetic field. For this purpose the variational 
approach is applied and the corresponding Lagrangian is written. 
\head
2. A quantum particle in the classical electromagnetic field. 
\endhead
    Massive particles in quantum mechanics are described by their wave functions. 
Various wave equations are usually written for these wave functions. The time 
dependent Shr\"odinger equation is one of them. It is written as follows:
$$
\hskip -2em
i\,\hbar\,\frac{\partial\psi}{\partial t}=
-\frac{\hbar^{\kern 1pt 2}}{2\,m}\nabla^{\kern 1pt 2}\psi.
\mytag{2.1}
$$ 
The equation \mythetag{2.1} can be derived variationally, using the following
action integral:
$$
\hskip -2em
S_\psi=\frac{i\,\hbar}{2}\int\Bigl(\frac{\partial\,\psi}{\partial\,t}\,\overline{\psi}
-\psi\,\frac{\partial\,\overline{\psi}}{\partial\,t}\,\Bigr)
\,d^{\kern 1pt 3}r\,d\kern 1pt t
-\frac{\hbar^{\kern 1pt 2}}{2\,m}\int |\nabla\psi|^2\,d^{\kern 1pt 3}r\,
d\kern 1pt t.
\mytag{2.2}
$$ 
Though the equation \mythetag{2.1} is not a relativistic equation, the 
integrals in \mythetag{2.2} are taken over the four-dimensional Minkowski 
space. The electromagnetic field is introduced into the equation \mythetag{2.1} 
through the following momentum transformation: 
$$
\xalignat 2
&\hskip -2em
\frac{i\,\hbar}{c}\,\frac{\partial}{\partial\,t}
\longrightarrow \frac{i\,\hbar}{c}\,\frac{\partial}{\partial\,t}
-\frac{e}{c}\,\phi,
&&-i\,\hbar\,\nabla \longrightarrow-i\,\hbar\,\nabla-\frac{e}{c}\,\bold A.
\quad
\mytag{2.3}
\endxalignat
$$
Here $\phi$ and $\bold A$ are the scalar potential and the vector potential of 
the electromagnetic field respectively. The transformation \mythetag{2.3} is
known as the minimal coupling\linebreak (see \mycite{3}). Applying the minimal 
coupling transformation \mythetag{2.3} to \mythetag{2.1}, we derive 
$$
i\,\hbar\,\frac{\partial\psi}{\partial t}=
-\frac{\hbar^{\kern 1pt 2}}{2\,m}\nabla^{\kern 1pt 2}\psi
+e\,\phi\,\psi+\frac{i\,e\,\hbar}{2\,m\,c}\bigl((\bold A,\nabla)
+(\nabla,\bold A)\bigr)\psi
+\frac{e^{\kern 1pt 2}\,|\bold A|^2}{2\,m\,\,c^{\kern 1pt 2}}\,\psi.
\quad
\mytag{2.4}
$$ 
The equation \mythetag{2.4} describes a non-relativistic quantum particle with 
the mass $m$ and with the electric charge $e$ in the classical electromagnetic 
field. Applying \mythetag{2.3} to \mythetag{2.2}, one can derive the action 
integral for the equation \mythetag{2.4}:
$$
\hskip -2em
\gathered
S_\psi=\frac{i\,\hbar}{2}\int\Bigl(\frac{\partial\,\psi}{\partial\,t}\,\overline{\psi}
-\psi\,\frac{\partial\,\overline{\psi}}{\partial\,t}\,\Bigr)
\,d^{\kern 1pt 3}r\,d\kern 1pt t
-e\!\int\!\!\phi\,|\psi|^2\,d^{\kern 1pt 3}r\,d\kern 1pt t\,-\\
-\,\frac{\hbar^{\kern 1pt 2}}{2\,m}\int |\nabla\psi|^2\,d^{\kern 1pt 3}r
\,d\kern 1pt t
-\frac{i\,e\,\hbar}{2\,m\,c}\int \overline{\psi}\,(\bold A,\nabla\psi)
\,d^{\kern 1pt 3}r\,d\kern 1pt t\,+\\
+\,\frac{i\,e\,\hbar}{2\,m\,c}\int
\psi\,(\bold A,\nabla\overline{\psi}\,)\,d^{\kern 1pt 3}r\,d\kern 1pt t
-\frac{e^2}{2\,m\,c^{\kern 1pt 2}}\int|\bold A|^2\,|\psi|^2\,d^{\kern 1pt 3}r
\,d\kern 1pt t.
\endgathered
\mytag{2.5}
$$
The action integral \mythetag{2.5} corresponds to a quantum particle 
experiencing the influence of the classical electromagnetic field. In order
to describe the electromagnetic field itself the action integral of the free
electromagnetic field is used (see \mycite{4} or \mycite{5}):
$$
\hskip -2em
S_{\text{fef}}=-\frac{1}{16\,\pi\,c}\int\sum^3_{p=0}
\sum^3_{q=0}F_{pq}\,F^{pq}\,d^{\kern 1pt 4}r=
\int\frac{|\bold E|^2-|\bold H|^2}{8\,\pi}\,d^{\kern 1pt 3}r\,d\kern 1pt t.
\mytag{2.6}
$$
If we need to describe the backward influence of a quantum particle upon 
the electromagnetic field, we should add the action integrals
\mythetag{2.5} and \mythetag{2.6}:
$$
\hskip -2em
S=S_\psi+S_{\text{fef}}
\mytag{2.7}
$$
The total action integral $S$ implies three Euler-Lagrange equations:
$$
\xalignat 3
&\hskip -2em
\frac{\delta S}{\delta\,\overline{\psi}}=0,
&&\frac{\delta S}{\delta\,\phi}=0,
&&\frac{\delta S}{\delta\,\bold A}=0.
\mytag{2.8}
\endxalignat
$$
The first equation \mythetag{2.8} leads to \mythetag{2.4}. The other two
equations \mythetag{2.8} coincide with the Maxwell equations comprising 
charges and currents:
$$
\xalignat 2
&\hskip -2em
\divr\bold E=4\,\pi\,\rho,
&&\rot\bold H-\frac{1}{c}\frac{\partial\,\bold E}{\partial\,t}
=\frac{4\,\pi}{c}\,\bold j.
\mytag{2.9}
\endxalignat
$$
Due to \mythetag{2.7} the charge density $\rho$ and the current density 
$\bold j$ in \mythetag{2.9} are determined by the action integral 
\mythetag{2.5} according to the formulas
$$
\xalignat 2
&\hskip -2em
\rho=-\frac{\delta S_\psi}{\delta\,\phi},
&&\bold j=c\ \frac{\delta S_\psi}{\delta\,\bold A}.
\mytag{2.10}
\endxalignat
$$
Substituting the explicit expression \mythetag{2.5} for $S_\psi$ into 
\mythetag{2.10}, we derive:
$$
\xalignat 2
&\hskip -2em
\rho=e\,|\psi|^2,
&&\bold j=\frac{i\,e\,\hbar}{2\,m}\,\bigl(\psi\,\nabla\overline{\psi}-
\overline{\psi}\,\nabla\psi\bigr)-\frac{e^{\kern 1pt 2}\bold A}
{m\,c}\,|\psi|^2.
\mytag{2.11}
\endxalignat
$$
The formulas \mythetag{2.11} are compatible with the probabilistic 
interpretations of the wave function $\psi$ (see \mycite{1} and
\mycite{2}).\par
     The following formulas are well known (see \mycite{4} or \mycite{5}):
$$
\xalignat 2
&\hskip -2em
\bold E=-\nabla\phi-\frac{1}{c}\,\frac{\partial\bold A}{\partial\,t},
&&\bold H=\rot\bold A.
\mytag{2.12}
\endxalignat
$$
They express the electric field $\bold E$ and the magnetic field $\bold H$
through the corresponding vectorial potential $\bold A$ and scalar potential
$\phi$.\par
\head
3. A couple of quantum particles\\ in the classical electromagnetic field. 
\endhead
    The hydrogen atom, as well as our present model, is a couple of two 
quantum particles --- the proton and the electron. Such a couple of particles 
is described by a single wave function $\Psi$ but depending on two spatial 
variables $\bold r_p$ and $\bold r_e$ which are radius-vectors of the proton 
and the electron respectively. Apart from $\bold r_p$ and $\bold r_e$, the joint 
wave function of the proton and the electron depends on the time variable:
$$
\hskip -2em
\Psi=\Psi(t,\bold r_p,\bold r_e).
\mytag{3.1}
$$
The analog of the action integral \mythetag{2.2} for the wave function 
\mythetag{3.1} is written as 
$$
\gathered
S_\Psi=\frac{i\,\hbar}{2}\int\Bigl(\frac{\partial\,\Psi}{\partial\,t}
\,\overline{\Psi}-\Psi\,\frac{\partial\,\overline{\Psi}}{\partial\,t}
\,\Bigr)\,d^{\kern 1pt 3}r_{\kern -1pt p}\,d^{\kern 1pt 3}r_{\kern -1pt e}
\,d\kern 1pt t\,-\\
\vspace{1ex}
-\int\biggl(\frac{\hbar^{\kern 1pt 2}\,|\nabla_{\kern -1pt p}\Psi|^2}{2\,m_p}
+\frac{\hbar^{\kern 1pt 2}\,|\nabla_{\kern -1pt e}\Psi|^2}{2\,m_e}\biggr)
d^{\kern 1pt 3}r_{\kern -1pt p}\,d^{\kern 1pt 3}r_{\kern -1pt e}\,d\kern 1pt t.
\endgathered
$$ 
Similarly, the analog of the action integral \mythetag{2.5} is written as 
follows:
$$
\hskip -2em
\gathered
S_\Psi=\frac{i\,\hbar}{2}\int\Bigl(\frac{\partial\,\Psi}{\partial\,t}\,\overline{\Psi}
-\Psi\,\frac{\partial\,\overline{\Psi}}{\partial\,t}\,\Bigr)
\,d^{\kern 1pt 3}r_{\kern -1pt p}\,d^{\kern 1pt 3}r_{\kern -1pt e}\,d\kern 1pt t\,-\\
\vspace{1ex}
-\,e\int\!\bigl(\phi_p-\phi_e\bigr)\,|\Psi|^2\,
d^{\kern 1pt 3}r_{\kern -1pt p}\,d^{\kern 1pt 3}r_{\kern -1pt e}\,d\kern 1pt t\,-\\
\vspace{1ex}
-\int\biggl(\frac{\hbar^{\kern 1pt 2}\,|\nabla_{\kern -1pt p}\Psi|^2}{2\,m_p}
+\frac{\hbar^{\kern 1pt 2}\,|\nabla_{\kern -1pt e}\Psi|^2}{2\,m_e}\biggr)
d^{\kern 1pt 3}r_{\kern -1pt p}\,d^{\kern 1pt 3}r_{\kern -1pt e}\,d\kern 1pt t\,-\\
\vspace{1ex}
-\,\frac{i\,e\,\hbar}{2\,c}\int\biggl(\frac{\overline{\Psi}\,(\bold A_p,
\nabla_{\kern -1pt p}\Psi)}{m_p}
-\frac{\overline{\Psi}\,(\bold A_e,\nabla_{\kern -1pt e}\Psi)}{m_e}\biggr)
\,d^{\kern 1pt 3}r_{\kern -1pt p}\,d^{\kern 1pt 3}r_{\kern -1pt e}\,d\kern 1pt t\,+\\
\vspace{1ex}
+\,\frac{i\,e\,\hbar}{2\,c}\int\biggl(\frac{\Psi\,(\bold A_p,\nabla_{\kern -1pt p}\overline{\Psi}\,)}{m_p}-\frac{\Psi\,(\bold A_e,\nabla_{\kern -1pt e}
\overline{\Psi}\,)}{m_e}\biggr)
\,d^{\kern 1pt 3}r_{\kern -1pt p}\,d^{\kern 1pt 3}r_{\kern -1pt e}\,d\kern 1pt t\,-\\
\vspace{1ex}
-\frac{e^2}{2\,c^{\kern 1pt 2}}\int\biggl(\frac{|\bold A_p|^2}{m_p}
+\frac{|\bold A_e|^2}{m_e}\biggr)\,|\Psi|^2
\,d^{\kern 1pt 3}r_{\kern -1pt p}\,d^{\kern 1pt 3}r_{\kern -1pt e}\,d\kern 1pt t.
\endgathered
\mytag{3.2}
$$ 
For the sake of brevity in \mythetag{3.2} the following notations are used:
$$
\xalignat 2
&\hskip -2em
\phi_p=\phi(t,\bold r_p), 
&&\bold A_p=\bold A(t,\bold r_p),\\
\vspace{-1.7ex}
\mytag{3.3}\\
\vspace{-1.7ex}
&\hskip -2em
\phi_e=\phi(t,\bold r_e), 
&&\bold A_e=\bold A(t,\bold r_e).
\endxalignat 
$$
Like in \mythetag{2.5}, in order to describe the compete system including the 
classical electromagnetic field we add the action integral \mythetag{2.6} to
\mythetag{3.2}: 
$$
\hskip -2em
S=S_\Psi+S_{\text{fef}}
\mytag{3.4}
$$
Then, like in \mythetag{2.8}, three Euler-Lagrange equations for $S$
in \mythetag{3.4} are written:
$$
\xalignat 3
&\hskip -2em
\frac{\delta S}{\delta\,\overline{\Psi}}=0,
&&\frac{\delta S}{\delta\,\phi}=0,
&&\frac{\delta S}{\delta\,\bold A}=0.
\mytag{3.5}
\endxalignat
$$\par
     The first equation \mythetag{3.5} leads to the time dependent 
Shr\"odinger equation for the proton and the electron in the classical
electromagnetic field. This equation is similar to the equation 
\mythetag{2.4}, but a little bit more complicated than it:
$$
\gathered
i\,\hbar\,\frac{\partial\Psi}{\partial t}=
-\frac{\hbar^{\kern 1pt 2}\,\nabla_{\kern -1pt p}^{\kern 1pt 2}\Psi}{2\,m_p}
-\frac{\hbar^{\kern 1pt 2}\,\nabla_{\kern -1pt e}^{\kern 1pt 2}\Psi}{2\,m_e}
+e\,\phi_p\,\Psi-e\,\phi_e\,\Psi\,+\\
+\,\frac{i\,e\,\hbar}{2\,m_p\,c}\bigl((\bold A_p,\nabla_{\kern -1pt p})
+(\nabla_{\kern -1pt p},\bold A_p)\bigr)\Psi
-\frac{i\,e\,\hbar}{2\,m_e\,c}\bigl((\bold A_e,\nabla_{\kern -1pt e})
+(\nabla_{\kern -1pt e},\bold A_e)\bigr)\Psi\,+\\
+\frac{e^{\kern 1pt 2}\,|\bold A_p|^2}{2\,m_p\,\,c^{\kern 1pt 2}}\,\Psi
+\frac{e^{\kern 1pt 2}\,|\bold A_e|^2}{2\,m_e\,\,c^{\kern 1pt 2}}\,\Psi.
\endgathered
\quad
\mytag{3.6}
$$ 
In the equation \mythetag{3.6} the same notations \mythetag{3.3} are used as in 
\mythetag{3.2}.\par
     The second and the third equations \mythetag{3.5} lead to the Maxwell 
equations \mythetag{2.9}. The formulas for $\rho$ and $\bold j$ are similar
to \mythetag{2.10}: 
$$
\xalignat 2
&\rho=-\frac{\delta S_\Psi}{\delta\,\phi},
&&\bold j=c\ \frac{\delta S_\Psi}{\delta\,\bold A}.
\endxalignat
$$
However, the ultimate expressions for them are different from \mythetag{2.11}:
$$
\hskip -2em
\gathered
\rho=e\int |\Psi_p|^2\,d^{\kern 1pt 3}r_{\kern -1pt e}
-e\int |\Psi_e|^2\,d^{\kern 1pt 3}r_{\kern -1pt p},\\
\vspace{2ex}
\gathered
\bold j=\frac{i\,e\,\hbar}{2\,m_p}\int\bigl(\Psi_p\,\nabla\overline{\Psi_p}-
\overline{\Psi_p}\,\nabla\Psi_p\bigr)\,d^{\kern 1pt 3}r_{\kern -1pt e}
-\frac{e^{\kern 1pt 2}\bold A}{m_p\,c}\int|\Psi_p|^2
\,d^{\kern 1pt 3}r_{\kern -1pt e}\,-\\
-\frac{i\,e\,\hbar}{2\,m_e}\int\bigl(\Psi_e\,\nabla\overline{\Psi_e}-
\overline{\Psi_e}\,\nabla\Psi_e\bigr)\,d^{\kern 1pt 3}r_{\kern -1pt p}
-\frac{e^{\kern 1pt 2}\bold A}{m_e\,c}\int|\Psi_e|^2
\,d^{\kern 1pt 3}r_{\kern -1pt p}.
\endgathered
\endgathered
\mytag{3.7}
$$
In the above formulas \mythetag{3.7} the following notations are used:
$$
\xalignat 2
&\hskip -2em
\Psi_p=\Psi(t,\bold r,\bold r_e),
&&\Psi_e=\Psi(t,\bold r_p,\bold r).
\mytag{3.8}
\endxalignat
$$
The notations \mythetag{3.8} are similar to the notations \mythetag{3.3}.
For the sake of completeness, in addition to \mythetag{3.6} and \mythetag{2.9},
one should write the equations \mythetag{2.12} expressing $\bold H$ and $\bold E$
through $\phi$ and $\bold A$.
\head
4. Spherically symmetric eigenstates of the model.
\endhead
     It is known that the ground state of the hydrogen atom is spherically 
symmetric. By analogy we look for eigenstates of our model in the class of
spherically symmetric wave functions. Each eigenstate is a stationary state.
Its time dependence is given by an oscillating exponential factor. As a result 
we get
$$
\hskip -2em
\Psi(t,\bold r_p,\bold r_e)=\psi(r_{\kern -1pt p},r_{\kern -1pt e})
\,\exp(-i\,\Cal E\,t/\hbar). 
\mytag{4.1}
$$
where $r_{\kern -1pt p}=|\bold r_p|$ and $r_{\kern -1pt e}=|\bold r_e|$. 
Apart from \mythetag{4.1}, we assume the electric and magnetic fields $\bold E$ 
and $\bold H$ associated with the eigenstate to be stationary and spherically 
symmetric. As a result, denoting $r=|\bold r|$, we get
$$
\xalignat 2
&\hskip -2em
\bold E(t,\bold r)=E(r)\,\frac{\bold r}{r},
&&\bold H(t,\bold r)=H(r)\,\frac{\bold r}{r}.
\mytag{4.2}
\endxalignat
$$
\mylemma{4.1} Each spherically symmetric stationary magnetic field given by the 
second formula \mythetag{4.2} is identically zero. 
\endproclaim
\demo{Proof} Note that the second equality \mythetag{2.12} is equivalent to the 
Maxwell equation $\divr\bold H=0$. The integral presentation of $\divr\bold H=0$
is 
$$
\hskip -2em
\oint\limits_{\!\!S}(\bold H,\,\bold n)\,dS=0, 
\mytag{4.3}
$$
where $S$ is an arbitrary closed surface and $\bold n$ is the unit normal vector of 
$S$. Substituting the second formula \mythetag{4.2} into \mythetag{4.3} and applying
\mythetag{4.3} to the sphere with the radius $r$ whose center is at the origin, we
derive
$$
\hskip -2em
\oint\limits_{\!\!S}(\bold H,\,\bold n)\,dS=4\,\pi\,r^2\,H(r)=0. 
\mytag{4.4}
$$
The equality \mythetag{4.4} yields $H(r)=0$. Thus Lemma~\mythelemma{4.1} is 
proved. 
\qed\enddemo
     Since $\bold H=0$, we can choose $\bold A=0$ for the corresponding vector
potential $\bold A$. The spherically symmetric stationary electric field given
by the first formula \mythetag{4.2} is associated with a spherically symmetric 
stationary scalar potential. Thus we have 
$$
\xalignat 2
&\hskip -2em
\phi(t,\bold r)=\phi(r), 
&&\bold A(t,\bold r)=0.
\mytag{4.5}
\endxalignat
$$
The next step is to apply \mythetag{2.12} and \mythetag{2.9} to \mythetag{4.5}. 
This yields
$$
\xalignat 2
&\hskip -2em
\frac{d^{\kern 0.5pt 2}\phi}{d\kern 1.5pt r^2}+
\frac{2}{r}\,\frac{d\,\phi}{d\kern 1.5pt r}=-4\,\pi\,\rho, 
&&\bold j=0.
\mytag{4.6}
\endxalignat
$$
The first equation \mythetag{4.6} is a version of the Poisson equation
$\triangle\phi=-4\,\pi\,\rho$ (see \mycite{6}).\par
     Now let's substitute \mythetag{4.6} along with \mythetag{4.1} into the
time-dependent Shr\"odinger equation \mythetag{3.6}. As a result this equation 
transforms to
$$
\hskip -2em
-\frac{\hbar^{\kern 1pt 2}\,\nabla_{\kern -1pt p}^{\kern 1pt 2}\psi}{2\,m_p}
-\frac{\hbar^{\kern 1pt 2}\,\nabla_{\kern -1pt e}^{\kern 1pt 2}\psi}{2\,m_e}
+e\,\phi_p\,\psi-e\,\phi_e\,\psi=\Cal E\,\psi.
\mytag{4.7}
$$ 
The notations \mythetag{3.3} now yield $\phi_p=\phi(r_{\kern -1pt p})$ and 
$\phi_e=\phi(r_{\kern -1pt e})$, where $r_{\kern -1pt p}=|\bold r_p|$ and
$r_{\kern -1pt e}=|\bold r_e|$. Keeping in mind these notations, we can 
separate variables and break \mythetag{4.7} into two equations if we 
substitute $\psi(r_{\kern -1pt p},r_{\kern -1pt e})=\psi_p(r_{\kern -1pt p})
\,\psi_e(r_{\kern -1pt e})$:
$$
\hskip -2em
\aligned
-\frac{\hbar^{\kern 1pt 2}\,\nabla_{\kern -1pt p}^{\kern 1pt 2}\psi_p}{2\,m_p}
+e\,\phi_p\,\psi_p=\Cal E_p\,\psi_p,\\
\vspace{1ex}
-\frac{\hbar^{\kern 1pt 2}\,\nabla_{\kern -1pt e}^{\kern 1pt 2}\psi_e}{2\,m_e}
-e\,\phi_e\,\psi_e=\Cal E_e\,\psi_e.
\endaligned
\mytag{4.8}
$$ 
The total eigenvalue $\Cal E$ is the sum of two eigenvalues $\Cal E_p$ and
$\Cal E_e$ in \mythetag{4.8}:
$$
\hskip -2em
\Cal E=\Cal E_p+\Cal E_e.
\mytag{4.9}
$$
     Having subdivided the equation \mythetag{4.7} into two equations \mythetag{4.8},
we can omit the particle indices in the arguments of the functions $\psi_p$ and
$\psi_e$, i\.\,e\. we can write $\psi_p=\psi_p(r)$ and $\psi_e=\psi_e(r)$, where
$r=|\bold r|$. Then the equations \mythetag{4.8} are written as 
$$
\hskip -2em
\aligned
-\frac{\hbar^{\kern 1pt 2}\,\nabla^{\kern 1pt 2}\psi_p}{2\,m_p}
+e\,\phi(r)\,\psi_p=\Cal E_p\,\psi_p,\\
\vspace{1ex}
-\frac{\hbar^{\kern 1pt 2}\,\nabla^{\kern 1pt 2}\psi_e}{2\,m_e}
-e\,\phi(r)\,\psi_e=\Cal E_e\,\psi_e.
\endaligned
\mytag{4.10}
$$ 
The functions $\psi_p=\psi_p(r)$ and $\psi_e=\psi_e(r)$ are spherically symmetric.
Therefore the equations \mythetag{4.10} are written as ordinary differential 
equations
$$
\hskip -2em
\aligned
-\frac{\hbar^{\kern 1pt 2}}{2\,m_p}\biggl(\frac{d^{\kern 1pt 2}\psi_p}
{d\kern 1.5pt r^2}+\frac{2}{r}\,\frac{d\,\psi_p}{d\kern 1.5pt r}\biggr)
+e\,\phi(r)\,\psi_p=\Cal E_p\,\psi_p,\\
\vspace{1ex}
-\frac{\hbar^{\kern 1pt 2}}{2\,m_e}\biggl(\frac{d^{\kern 1pt 2}\psi_e}
{d\kern 1.5pt r^2}+\frac{2}{r}\,\frac{d\,\psi_e}{d\kern 1.5pt r}\biggr)
-e\,\phi(r)\,\psi_e=\Cal E_e\,\psi_e.
\endaligned
\mytag{4.11}
$$\par 
     Though the equations \mythetag{4.11} look like two independent equations,
they are not actually independent since $\phi(r)$ is not a given 
function in them. It is defined by the first equation \mythetag{4.6}. In order 
to specify the charge density $\rho$ in the right hand side of this equation
we use the first formula \mythetag{3.7}. Upon substituting the product 
$\psi(r_{\kern -1pt p},r_{\kern -1pt e})
=\psi_p(r_{\kern -1pt p})\,\psi_e(r_{\kern -1pt e})$ and \mythetag{4.9} into 
\mythetag{4.1} we get 
$$
\hskip -2em
\Psi(t,\bold r_p,\bold r_e)=\psi_p(r_p)\,\psi_e(r_e)
\,\exp(-i\,\Cal E_p\,t/\hbar)\,\exp(-i\,\Cal E_e\,t/\hbar). 
\mytag{4.12}
$$
The wave functions $\psi_p=\psi_p(r)$ and $\psi_e=\psi_e(r)$ in \mythetag{4.12} 
are assumed to be normalized to unity according to their probabilistic 
interpretations:
$$
\hskip -2em
\aligned
\int\!|\psi_p|^2\,d^{\kern 1pt 3}r=4\,\pi\int\limits^{\,\infty}_{\!0}
\!|\psi_p(r)|^2\,r^2\,d\kern 1pt r=1,\\
\int\!|\psi_e|^2\,d^{\kern 1pt 3}r=4\,\pi\int\limits^{\,\infty}_{\!0}
\!|\psi_e(r)|^2\,r^2\,d\kern 1pt r=1.
\endaligned
\mytag{4.13}
$$
Substituting \mythetag{4.12} into the first formula \mythetag{3.7} and taking into
account \mythetag{4.13}, we derive the following expression for the charge density
$\rho$:
$$
\hskip -2em
\rho=\rho(r)=e\,|\psi_p(r)|^2-e\,|\psi_e(r)|^2.
\mytag{4.14}
$$
Combining \mythetag{4.14} with the first equation \mythetag{4.6}, we get
the differential equation
$$
\hskip -2em
\frac{d^{\kern 0.5pt 2}\phi}{d\kern 1.5pt r^2}+
\frac{2}{r}\,\frac{d\,\phi}{d\kern 1.5pt r}=4\,\pi\,e\,|\psi_e(r)|^2
-4\,\pi\,e\,|\psi_p(r)|^2.
\mytag{4.15}
$$
The equations \mythetag{4.11} complemented with the equation \mythetag{4.15}
constitute a closed system of three ordinary differential equations for two
complex-valued functions $\psi_p=\psi_p(r)$ and $\psi_e=\psi_e(r)$ and for
one real-valued function $\phi=\phi(r)$.\par
\mydefinition{4.1} A solution of the system of differential equations
\mythetag{4.11} and \mythetag{4.15} on the half-line $0<r<+\infty$ is called 
an {\it eigenstate\/} of the model if the functions $\psi_p=\psi_p(r)$ and 
$\psi_e=\psi_e(r)$ are square-integrable in the sense of the formulas 
\mythetag{4.13} along with their derivatives so that 
$$
\hskip -2em
\aligned
-\int\!\overline{\psi_p}\,\nabla^{\kern 1pt 2}\psi_p
\,d^{\kern 1pt 3}r=\int\!|\nabla\psi_p|^2\,d^{\kern 1pt 3}r<\infty,\\
-\int\!\overline{\psi_e}\,\nabla^{\kern 1pt 2}\psi_e
\,d^{\kern 1pt 3}r=\int\!|\nabla\psi_e|^2\,d^{\kern 1pt 3}r<\infty.
\endaligned
\mytag{4.16}
$$
\enddefinition
\mylemma{4.2} For any eigenstate of the model given by a solution of the equations 
\mythetag{4.11} and \mythetag{4.15} not only the total eigenvalue $\Cal E$ in 
\mythetag{4.9} is a real number, but the separate eigenvalues $\Cal E_p$ and 
$\Cal E_e$ are also real numbers. 
\endproclaim
     Lemma~\mythelemma{4.2} is easily proved by applying \mythetag{4.13} and
\mythetag{4.16} to the equations \mythetag{4.11} written in the form of
\mythetag{4.10}.\par
     Note that if the functions $\psi_p=\psi_p(r)$ and $\psi_e=\psi_e(r)$ are
given, the equation \mythetag{4.15} is a linear non-homogeneous ordinary 
differential equation \pagebreak of the second order. Its general solution 
is defined by the formula with two arbitrary constants 
$$
\hskip -2em
\phi=\phi_{\kern 1pt 0}+C_1+\frac{C_2}{r^2}, 
\mytag{4.17}
$$
provided some particular solution $\phi=\phi_{\kern 1pt 0}(r)$ is known. 
\mylemma{4.3} For any eigenstate of the model given by a solution of the equations 
\mythetag{4.11} and \mythetag{4.15} there is a particular solution of the equation
\mythetag{4.15} such that
$$
\hskip -2em
\phi_{\kern 1pt 0}(r)=\int\limits_{\!r}^{\,\infty}\!\frac{1}{r^2}\,\biggl(\,\,\int
\limits_{\!0}^{\,r}4\,\pi\,e\,\bigl(|\psi_p(r)|^2-|\psi_e(r)|^2\bigr)\,r^2
\,d\kern 1pt r\!\biggr)d\kern 1pt r. 
\mytag{4.18}
$$
\endproclaim
     The integrals in \mythetag{4.18} do exist and are finite since $\psi_p$ and 
$\psi_e$ are square-integrable functions in the sense of \mythetag{4.13}. Therefore
Lemma~\mythelemma{4.3} is easily proved by direct calculations upon substituting 
\mythetag{4.18} into \mythetag{4.15}. 
     Note that the function \mythetag{4.18} obeys the condition $\phi_{\kern 1pt 0}
\longrightarrow 0$ as $r\longrightarrow +\infty$. Combining this fact with the
formula \mythetag{4.17}, we derive the following lemma. 
\mylemma{4.4} For any eigenstate of the model given by a solution of the equations 
\mythetag{4.11} and \mythetag{4.15} there is a finite limit 
$C=\dsize\lim_{r\to+\infty}\phi(r)\neq\infty$.
\endproclaim
     Note that the equations \mythetag{4.11} and \mythetag{4.15} are invariant
under the transformations
$$
\xalignat 3
&\hskip -2em
\phi\to\phi-C,
&&\Cal E_p\to \Cal E_p-e\,C,
&&\Cal E_e\to \Cal E_e+e\,C,
\mytag{4.19}
\endxalignat
$$
where $C$ is some constant. The transformations \mythetag{4.19} do not 
change the total eigenvalue $\Cal E$ in \mythetag{4.9}.\par 
     Combining Lemma~\mythelemma{4.4} with \mythetag{4.19}, we can formulate the
next lemma. 
\mylemma{4.5} For any eigenstate of the model given by a solution of the equations 
\mythetag{4.11} and \mythetag{4.15} there is an associated eigenstate such that
$\dsize\lim_{r\to+\infty}\phi(r)=0$. The partial eigenvalues $\Cal E_p$ and 
$\Cal E_e$ of this associated eigenstate are called energy levels of the proton
and the electron respectively. 
\endproclaim
      If the function $\phi=\phi(r)$ is fixed, then the equations \mythetag{4.11}
are linear homogeneous ordinary differential equations of the second order with 
respect to the functions $\psi_p$ and $\psi_e$. These equations are very similar 
to each other. Therefore we can write them in a unified way omitting the particle 
indices:
$$
\hskip -2em
-\frac{\hbar^{\kern 1pt 2}}{2\,m}\biggl(\psi\,''
+\frac{2}{r}\ \psi\,'\biggr)
+u(r)\,\psi=\varepsilon\,\psi.
\mytag{4.20}
$$
The square-integrability conditions \mythetag{4.13} and \mythetag{4.16} can
also be written in a unified way as the following two conditions for solutions
of the equation \mythetag{4.20}:
$$
\xalignat 2
&\hskip -2em
\int\limits^{\,\infty}_{\!0}|\psi|^2\,r^2\,d\kern 1pt r<\infty,
&&\int\limits^{\,\infty}_{\!0}|\psi'|^2\,r^2\,d\kern 1pt r<\infty.
\mytag{4.21}
\endxalignat
$$\par
     For any two solutions $\psi_1$ and $\psi_2$ of the differential equation
\mythetag{4.20} we construct their Wronskian: $W=W(\psi_1,\psi_2)=\psi_1
\,\psi_2^{\,\prime}-\psi_1^{\,\prime}\,\psi_2$ (see \mycite{7}). One can 
easily prove the following facts concerning the Wronskian $W$:
\roster
\rosteritemwd=2pt
\item "1)" for any two solutions $\psi_1$ and $\psi_2$ of the differential equation
\mythetag{4.20} their Wronskian $W=W(\psi_1,\psi_2)$ obeys the first order 
ordinary differential equation $W\kern 1pt '+2\,W/r=0$ whose general solution is
$$
W=\frac{C}{r^2}\text{, \ where \ }C=\const;
\mytag{4.22}
$$
\item "2)" two solutions $\psi_1$ and $\psi_2$ of the differential equation
\mythetag{4.20} are linearly dependent if and only if their Wronskian 
$W=W(\psi_1,\psi_2)$ is zero. 
\endroster
\mylemma{4.6} Any two solutions $\psi_1$ and $\psi_2$ of the differential equation
\mythetag{4.20} obeying both square-integrability conditions \mythetag{4.21} are
linearly dependent. 
\endproclaim
\demo{Proof} It is known that the product of any two square-integrable functions 
is integrable (see \mycite{8}). Using this fact, from \mythetag{4.21} we derive
$$
\biggl|\,\int\limits^{\,\infty}_{\!0}\!W\,r^2\,d\kern 1pt r\biggr|\leqslant
\int\limits^{\,\infty}_{\!0}\!|\psi_1\,\psi_2'-\psi_1'\,\psi_2|\,r^2
\,d\kern 1pt r\leqslant
\int\limits^{\,\infty}_{\!0}(|\psi_1\,\psi_2'|+|\psi_1'\,\psi_2|)\,r^2
\,d\kern 1pt r<\infty.\quad
\mytag{4.23}
$$
On the other hand, substituting \mythetag{4.22} into the left hand side of 
\mythetag{4.23}, we get 
$$
\hskip -2em
\biggl|\,\int\limits^{\,\infty}_{\!0}\!W\,r^2\,d\kern 1pt r\biggr|
=\biggl|\,\int\limits^{\,\infty}_{\!0}C\,d\kern 1pt r\biggr|=
\cases 0 &\text{if\ \ } C=0,\\ \infty &\text{if\ \ } C\neq 0.
\endcases
\mytag{4.24}
$$
Comparing \mythetag{4.24} and \mythetag{4.23}, we find that $C=0$ and 
$W=W(\psi_1,\psi_2)=0$. The solutions $\psi_1$ and $\psi_2$ are linearly dependent
since their Wronskian is zero.
\qed\enddemo
\mylemma{4.7} Any nonzero complex-valued solution $\psi_{\sssize\Bbb C}$ of the 
differential equation \mythetag{4.20} obeying both square-integrability conditions \mythetag{4.21} is produced from some real-valued solution $\psi_{\sssize\Bbb R}$ 
of the same differential equation obeying both conditions \mythetag{4.21} by 
multiplying it by some complex constant: $\psi_{\sssize\Bbb C}
=C\,\psi_{\sssize\Bbb R}$.
\endproclaim
\demo{Proof} The equation \mythetag{4.20} is a linear ordinary differential equation 
with real coefficients. Therefore, if $\psi_{\sssize\Bbb C}$ is its nonzero solution, 
then the complex conjugate function $\overline{\psi_{\sssize\Bbb C}}$ is also a solution
of the equation \mythetag{4.20}. It is clear that if $\psi_{\sssize\Bbb C}$ obeys both
conditions \mythetag{4.21}, then so does the complex conjugate function $\overline{\psi_{\sssize\Bbb C}}$. Applying Lemma~\mythelemma{4.6} to the functions
$\psi_{\sssize\Bbb C}$ and $\overline{\psi_{\sssize\Bbb C}}$, we find that they are
linearly dependent, i\.\,e\. $\overline{\psi_{\sssize\Bbb C}}=K\,\psi_{\sssize\Bbb C}$,
where $K$ is some complex constant. Now we define two real-valued functions 
$$
\xalignat 2
&\hskip -2em
\psi_1=\frac{\psi_{\sssize\Bbb C}+\overline{\psi_{\sssize\Bbb C}}}{2}=
\frac{1+K}{2}\,\psi_{\sssize\Bbb C},
&&\psi_2=\frac{\psi_{\sssize\Bbb C}-\overline{\psi_{\sssize\Bbb C}}}{2\,i}=
\frac{1-K}{2\,i}\,\psi_{\sssize\Bbb C}.
\quad
\mytag{4.25}
\endxalignat
$$
It is easy to see that the functions \mythetag{4.25} are real-valued solutions 
of the equation \mythetag{4.20} obeying both square-integrability conditions 
\mythetag{4.21}. At least one of them is nonzero. Therefore at least one of 
the following two formulas 
$$
\pagebreak
\xalignat 2
&\psi_{\sssize\Bbb C}=\frac{2}{1+K}\,\psi_1,
&&\psi_{\sssize\Bbb C}=\frac{2\,i}{1-K}\,\psi_2
\endxalignat
$$
yields a valid expression $\psi_{\sssize\Bbb C}=C\,\psi_{\sssize\Bbb R}$
proving Lemma~\mythelemma{4.7}. 
\qed\enddemo
     Now we return to the equations \mythetag{4.11} and apply 
Lemma~\mythelemma{4.7} to them. As a result we easily prove the following 
lemma strengthening Lemma~\mythelemma{4.5}. 
\mylemma{4.8} For any eigenstate of the model given by a solution of the equations 
\mythetag{4.11} and \mythetag{4.15} there is an associated eigenstate with purely 
real-valued eigenfunctions $\psi_p(r)$ and $\psi_e(r)$ such that $\dsize\lim_{r\to+\infty}\phi(r)=0$. 
\endproclaim
     Let's recall that apart from the equations \mythetag{4.11} and \mythetag{4.15} 
we have the equality $\bold j=0$ in \mythetag{4.6}. Applying the second formula 
\mythetag{3.7} to it and taking into account the equality $\bold A=0$ from 
\mythetag{4.5}, we derive the equation 
$$
\hskip -2em
\frac{i\,e\,\hbar}{2\,m_p}\,\bigl(\psi_p\,\nabla\overline{\psi_p}-
\overline{\psi_p}\,\nabla\psi_p\bigr)=
\frac{i\,e\,\hbar}{2\,m_e}\,\bigl(\psi_e\,\nabla\overline{\psi_e}-
\overline{\psi_e}\,\nabla\psi_e\bigr).
\mytag{4.26}
$$
The equation \mythetag{4.26} is fulfilled identically 
if we apply Lemma~\mythelemma{4.8} and choose purely real-valued eigenfunctions 
$\psi_p$ and $\psi_e$. But even if they are not purely real-valued, according to
Lemma~\mythelemma{4.7}, they differ from purely real ones by some constant
factors. In this case the equation \mythetag{4.26} is also fulfilled identically. 
\par
     Going on, one can use more complicated methods from \mycite{9} for studying 
the equations \mythetag{4.11}. Conjecturing some integrability conditions for 
$\phi(r)$ at infinity, one can prove that the energy levels $\Cal E_p$ and $\Cal E_e$
are non-positive, i\.\,e\. $\Cal E_p\leqslant 0$ and $\Cal E_e\leqslant 0$. 
The Volterra-type integral equations from \mycite{9} can be applied for calculating 
$\psi_p(r)$, $\psi_e(r)$, and $\phi(r)$ numerically along with the associated 
eigenvalues $\Cal E_p$ and $\Cal E_e$. However these steps require much more efforts 
such as choosing a proper software and writing computer code. They are left for a 
separate paper. 
\head
5. Comparison with the standard model\\ of the hydrogen atom. 
\endhead
     As the reader can see above the standard model of the hydrogen atom is much
more simple than the present model. Its spectrum is calculated explicitly, while 
our model require numeric computations. But there is also a conceptual difference.
In the standard model the proton and the electron behave as classical point charges 
when producing the electromagnetic field in the form of Coulomb potential. Then 
they interact with the produced Coulomb field quantum mechanically as distributed
charges according to their wave function in the Shr\"odinger equation with the
Hamilton operator \mythetag{1.1}.\par
     Our present model is more logically coherent. The proton and the electron in
this model behave quantum mechanically as distributed sources in creating the 
electromagnetic field (see formulas \mythetag{3.7}) and then they behave again 
quantum mechanically when interacting with the created field (see Shr\"odinger 
equation \mythetag{3.6}).\par 
     Which of the two models is more congruent to the nature? I hope to answer 
this question upon computing the spectrum of the new model in forthcoming papers. 
\par

\enddocument
\end